\documentclass[twocolumn,prl,showpacs,nofootinbib,superscriptaddress]{revtex4}
\usepackage{latexsym}
\usepackage{dcolumn}
\usepackage{epsfig}

\def\spa{\sigma_{p{-}\rm air}}
\def\spai{\sigma_{p{-}\rm air}^{\rm prod}}

\newcommand{\ba}{\begin{eqnarray}}

\newcommand{\ea}{\end{eqnarray}}
\newcommand{\be}{\begin{equation}}
\newcommand{\ee}{\end{equation}}

\newcommand{\eq}[1]{Eq.\,(\ref{#1})}

\newcommand{\sigtot}{\sigma_{\rm tot}}

\def\bea{\begin{eqnarray}} 
\def\eea{\end{eqnarray}}

\begin{document}

\title{Ultra-high Energy Predictions of Proton-Air Cross Sections from Accelerator Data: an Update}

\author{M.~M.~Block}
\affiliation{Department of Physics and Astronomy, Northwestern University, 
Evanston, IL 60208}

\begin{abstract}
At  the $pp$ center of mass  energy  $\sqrt s = 57\pm 7$ TeV, the Pierre Auger Observatory  (PAO) collaboration has recently measured the  proton-air inelastic production 
 cross section $\spai$, using a cosmic ray beam consisting  mainly of protons, with some helium contamination. Assuming  a helium contamination of 25\%, they subtracted 30 mb from their measured $\spai$, resulting in a p-air inelastic production cross section, $\spai=475 \pm 22\  ({\rm  stat.})\pm^{20}_{15} \ ({\rm  syst.})$ mb, where (stat.) is the statistical error and (syst.) is the the systematic error, exclusive of helium contamination. Using this result in a Glauber calculation to obtain the $pp$ inelastic cross section, at 57 TeV they found the inelastic $pp$ cross section $\sigma_{\rm inel}= 90\pm 7\  ({\rm stat.}) \pm^9_{11} ({\rm syst.}) \pm 1.5 {\rm \  (Glaub.})$ mb, where (syst.) is the systematic and (Glaub.) is the error associated with the Glauber calculation needed to convert $\spai$ to pp $\sigma_{\rm inel}$.   Parameterization of the $\bar pp$ and $pp$ cross sections incorporating analyticity constraints and unitarity has allowed us to make accurate extrapolations to ultra-high energies, and, using Glauber calculations, also accurately predict cosmic ray results for  $\spai$. In this update for 57 TeV, we predict i)  a $pp$ total cross section, $\sigma_{\rm tot}=133.4\pm 1.6$ mb, using high energy predictions from a saturated Froissart bound parameterization of accelerator data on forward $\bar pp$ and $pp$ scattering amplitudes and ii) a p-air inelastic production cross section,  $\spai=483\pm 3 $ mb, by using $\sigma_{\rm tot}$ together with Glauber theory. Using the POA estimates of the variation of their measured $\spai$ with helium contamination, we were able to determine independently that the helium contamination was 19\%, in reasonable  agreement with their estimate of 25\%.  Our predictions agree with all available cosmic ray extensive air shower measurements, both in magnitude and in energy dependence. Further, by using our value for the $pp$ total cross section at 57 TeV, Block and Halzen \cite{blackdisk} have predicted that the $pp$ inelastic cross section is $\sigma_{\rm inel}= 92.9\pm 1.6$ mb, in  agreement with the measured POA value.
\end{abstract}


\maketitle

{\em Introduction.}
The PAO collaboration \cite{Auger} has recently published the value for the proton-air inelastic production cross section  $\spai=505\pm22\  ({\rm  stat.})\ {}^{+20}_{-15} \ ({\rm  syst.})$ mb, at $\sqrt s=57\pm7$ GeV, where $\sqrt s$ is the $pp$ center of mass (cms) energy, if the cosmic ray beam consists exclusively of protons. If the beam is contaminated by helium nuclei, the PAO collaboration \cite{Auger} states that the measured cross section is reduced by 0, -12, -30 and -80 mb for a 0, 10, 25 and 50\% contamination, respectively. They also measured \cite{Augerinelastic}  the inelastic $pp$ cross section $\sigma_{\rm inel}= 90\pm 7 ({\rm stat.}) \pm^9_{11} ({\rm syst.}) \pm 1.5 {\rm \  (Glaub.})$ mb, assuming a 25\% helium contamination. 

This note is a very short update
of Block's \cite{Blockcr} paper, ``Ultra-high Energy Predictions of protn-air Cross Secctions from Accelerator Data'', which has made very accurate predictions at cosmic ray energies for the total $pp$ cross section, $\sigma_{pp}$,  from fits \cite{BH} to accelerator data that used adaptive data sifting algorithms \cite{sieve} and analyticity constraints {\cite{blockanalyticity} that were not available in the earlier work of Block, Halzen and Stanev \cite{blockhalzenstanev}. Using  these results and the available cosmic ray measurements,  an excellent fit to the then-available cosmic ray $\spai$  measurements was made. In this note, we discuss our predictions for the $pp$ cms energy of 57 TeV at which the PAO results were obtained. 

The purpose of this update is to i)  make an {\em accurate} prediction of $\spai$, the 57 TeV cosmic ray p-air total cross section 
, and ii) from this prediction, estimate the helium contamination of the PAO result,  iii) predict the  57 TeV $\sigma_{\rm tot}$, the $pp$ total cross section  and finally, iv) compare our  prediction of $\sigma_{\rm inel}$,  the $pp$ inelastic cross section, with the PAO result \cite{Augerinelastic}.
This note is  purposely intended to be very brief; the relevant details of the fundamental calculations can be found in the original paper of Block \cite{Blockcr} and will not be presented here.  

{\em Determination of $\sigma_{pp}(s)$}
Block and Halzen \cite{BH} have made an analytic amplitude fit that saturates the Froissart bound \cite{froissart}, to data for both the high energy total cross section and the $\rho$-value, where $\rho$ is defined as the ratio of the real to the imaginary portion of the forward scattering amplitude, for both $\bar pp$ and $pp$ interactions.  Their high-energy behavior is parametrized using the analytic amplitude form
\ba
\sigma^{\pm}(\nu)&=&c_0+c_1\ln\left(\frac{\nu}{m}\right)+c_2\ln^2\left(\frac{\nu}{m}\right)+\beta_{\cal P'}\left(\frac{\nu}{m}\right)^{\mu -1}\nonumber\\
&&\pm\  \delta\left({\nu\over m}\right)^{\alpha -1},\label{sigmapmpp}\\
\rho^\pm(\nu)&=&{1\over\sigma^\pm(\nu)}\left\{\frac{\pi}{2}c_1+c_2\pi \ln\left(\frac{\nu}{m}\right)\right.\nonumber\\
&&\left.-\beta_{\cal P'}\cot({\pi\mu\over 2})\left(\frac{\nu}{m}\right)^{\mu -1}+\frac{4\pi}{\nu}f_+(0)\right.\nonumber\\
&&\left.\qquad\qquad\qquad\pm \delta\tan({\pi\alpha\over 2})\left({\nu\over m}\right)^{\alpha -1} \right\}\label{rhopmpp},
\ea
where the upper sign is for $pp$ and the lower for  $\bar pp$ scattering Here $\nu$ is the laboratory energy, $m$ the proton mass, $\mu=0.5$, and $f_+(0)$ is a dispersion relation subtraction constant. The 7 real constants $c_0,c_1,c_2,\beta_{\cal P'},\delta,\alpha$ and $f_+(0)$ are parameters of the fit. At high energies, $s$, the square of the cms  energy, approaches $2m\nu$; hence, we see from \eq{sigmapmpp}  that the cross sections behave as $\ln^2s$ at high energies, thus saturating the Froissart bound \cite{froissart}. From \eq{rhopmpp}, we see that $\rho\rightarrow 0$ as $1/\ln s$ as $s\rightarrow\infty$. 
    
Using 4 analyticity constraints\cite{blockanalyticity}, resulting from finite energy sum rules that used very high accuracy low-energy cross section measurements ($2\le \sqrt s\le 4$ GeV), they anchored both the cross sections $\sigma_{\bar pp}$ and $\sigma_{pp}$ and their laboratory energy derivatives to data at $\sqrt s=4$ GeV, thus reducing the number of parameters to from 7 to 4. The fit was to data with $6\le \sqrt s\le 1800$ GeV. This use of analyticity constraints  resulted in an excellent fit that, in turn, constrained $pp$ cross sections at cosmic ray energies to an accuracy $\sim 1-2$\%, even though (conflicting) Tevatron data provided the highest energy input. The Block and Halzen  fits  \cite {BH} to the $pp$ and $\bar pp$ cross sections are shown in Fig. \ref{fig:pp}. The PAO  energy of 57 TeV is indicated by the dotted line crossing  the fit. Our predicted $pp$ total cross section at 57 TeV is $\sigma_{\rm tot}=133.4\pm 1.6$ mb  For brevity, we have not shown the fit to $\rho$; see Block \cite{Blockcr}.
  
\begin{figure}[h,t,b] 
\begin{center}
\mbox{\epsfig{file=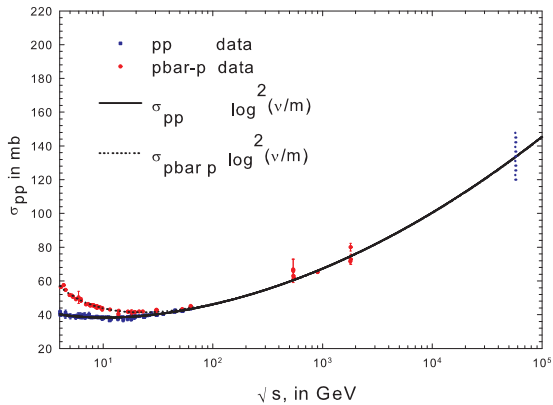
,width=3in%
,bbllx=26pt,bblly=325pt,bburx=565pt,bbury=735pt,clip=%
}}
\end{center}
\caption[]{
The fitted total cross section, $\sigtot$,  for $\bar pp$ (dashed curve)  and $pp$ (dot-dashed curve)  from \eq{sigmapmpp}, in mb vs. $\sqrt s$, the cms energy in GeV, taken from BH  \cite{BH}. The $\bar pp$ data used in the fit  are the (red) circles and the $pp$ data are the (blue) squares. The fitted data were anchored by values of $\sigtot^{\bar pp}$ and $\sigtot^{pp}$, together with the energy derivatives  ${d\sigtot^{\bar pp}/ d\nu}$ and ${d\sigtot^{pp}/ d\nu}$ at 6 GeV using FESR, as described in Ref. \cite{BH}. The vertical dotted line at 57000 GeV that intercepts the fit indicates the $pp$ cms of the PAO cosmic ray experiment \cite{Auger}.
\label{fig:pp}
}
\end{figure}

{\em Comparison of $\spa$ with cosmic ray data.}
\begin{figure}[h]
\begin{center}
\mbox{\epsfig{file=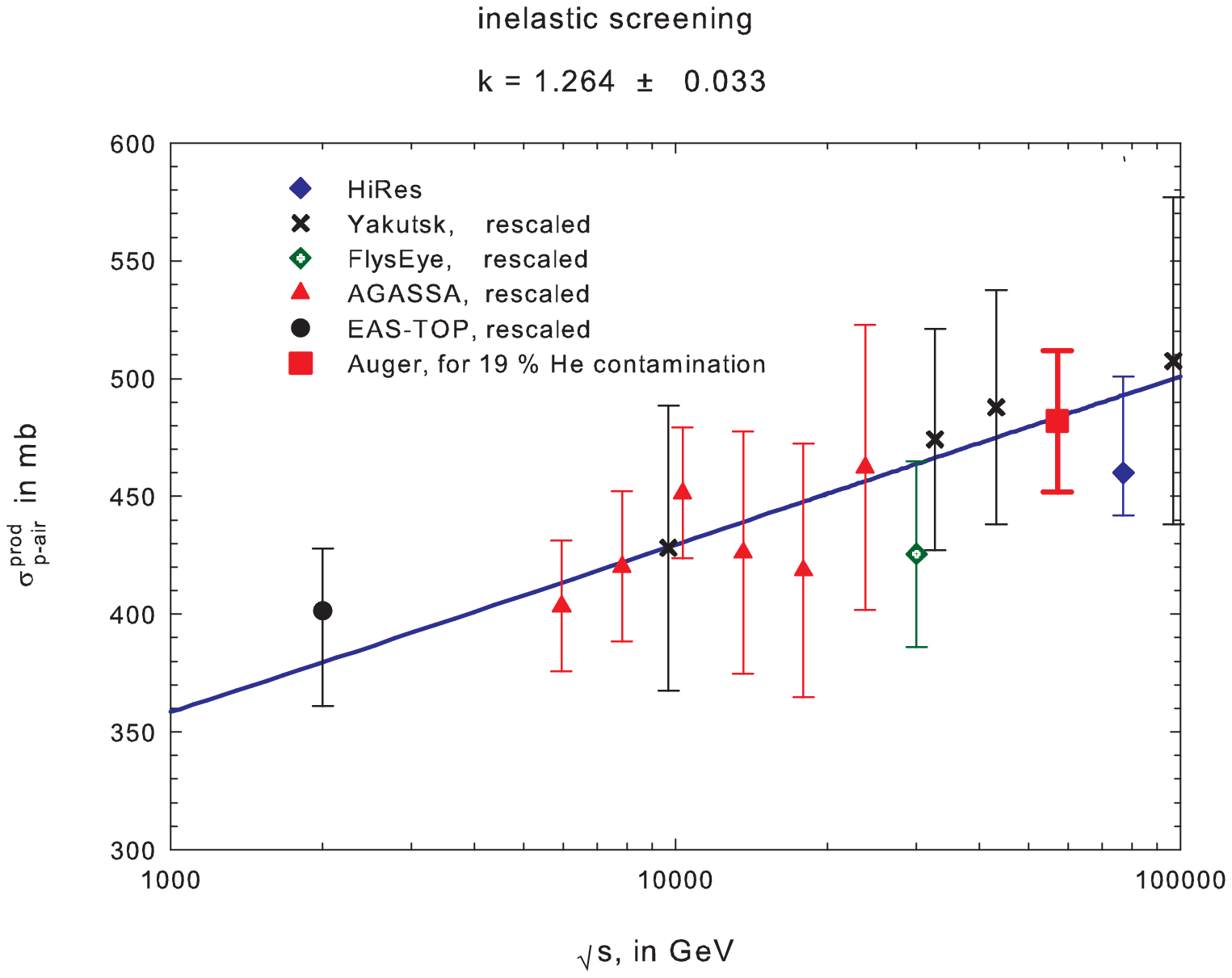
,width=3in%
,bbllx=122pt,bblly=331pt,bburx=570pt,bbury=634pt,clip=%
}}
\end{center}
\caption[]{\protect
{  A $\chi^2$ fit of the  renormalized AGASA, EASTOP, Fly's Eye and Yakutsk data for $\spai$, in mb,
 as a function of the energy, $\sqrt s$, in GeV. The HiRes point
 (solid diamond), at $\sqrt s=57000$ GeV, is model-independent and has  not been renormalized.  The renormalized ARGO-YBJ data were not used in the fit. Details of the renormalization are given in Block \cite{Blockcr}, as are references to the cosmic ray data shown in the Figure. The large (red) square is the Auger  measurement, adjusted for a 19\% helium contamination (see text). 
}
}
\label{fig:p-air}
\end{figure}
 In Fig. \ref{fig:p-air} we have plotted all available cosmic ray data for $\spai$, the proton-air  inelastic production cross section, in mb, as a function of $\sqrt s$, in GeV. The $\spai$ curve, derived from the total cross section fit of \eq{sigmapmpp} and utilizing the elastic slope parameter $B\equiv d \left[\ln d\sigma_{\rm el}/dt\right]|_{t=0}$ in a Glauber calculation, is taken from Block \cite{Blockcr}. All of the cosmic ray data but  the HiRes point were renormalized using a $k$-factor (k=1.264).  For brevity, we omit the discussion of this renormalization procedure as well as detailed references to the cosmic data used in the Figure; for complete information,  see Block \cite{Blockcr}. 

The  new point added in Fig. \ref{fig:p-air}} is the large (red) square, whose central value is our prediction of $\spai$ for 57 TeV and whose error \cite{Auger} is the total PAO experimental error excluding beam contamination uncertainty.  As we will show in the next Section, this  value, $\spai=482\pm 30$ mb, has a central value that corresponds to a 19\% helium contamination.  

\begin{figure}[h]
\begin{center}
\mbox{\epsfig{file=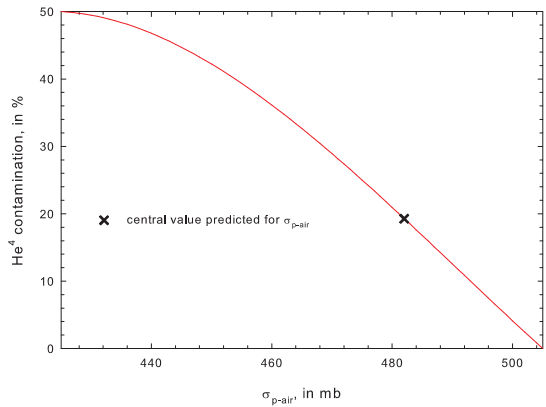
,width=3in%
,bbllx=75pt,bblly=240pt,bburx=505pt,bbury=550pt,clip=%
}}
\end{center}
\caption[ ]{\protect {Helium contamination of the  cosmic ray beam, in \%, vs. $\spai$, the  inelastic p-air cross section, in mb, for the PAO   experiment \cite{Auger}. The large cross  is our prediction for $\spai$, corresponding  to a 19\% helium contamination (see text). 
}
}
\label{fig:He}
\end{figure}
{\em Determination of the helium contamination in the cosmic ray `proton' air showers.}
 The PAO collaboration \cite{Auger} notes that ``We
recognise (sic) and identify the unknown mass composition of
cosmic rays as the major source of systematic uncertainty
for the proton-air cross-section analysis and we evaluate
its impact on the final result.'' They estimated that their best value for helium contamination was 25\%.  In this Section, we  obtain an  independent confirmation of this estimate.

We plot in Fig. \ref{fig:He} the helium fraction of the cosmic ray beam, in \%, vs. the PAO collaboration's  \cite{Auger} corrected measurements for $\spai$, in mb.  The large cross marks the point on the curve that is our prediction for the  p-air inelastic production cross section, $\spai= 482$ mb, which corresponds to a $19\pm 1\%$ helium contamination, in qualitative agreement with the Auger estimate of 25\%. 

{\em Prediction of the  $pp$ inelastic cross section, $\sigma_{\rm inel}$ at $\sqrt s =57$ TeV.}
 As mentioned earlier, the PAO collaboration \cite{Augerinelastic} also measured  the inelastic $pp$ scattering at 57 TeV.  By using a Glauber calculation, they converted  their measured value of $\spai$ (after allowing for   a 25 \% helium contamination) into a $pp$ inelastic cross section. They found $\sigma_{\rm inel}= 90\pm 7\  ({\rm stat.}) \pm^9_{11} \ ({\rm syst.}) \pm 1.5 {\rm \  (Glaub.})$ mb. 

Using our predicted $pp$ total cross section $\sigma_{\rm tot} =133.4\pm1.6$ mb, Block and Halzen \cite{blackdisk} determined that the $pp$ inelastic cross section at 57 TeV was given by $\sigma_{\rm inel}= 92.9\pm 1.6$ mb, which is in excellent agreement with the PAO value.

{\em Conclusions.}
 At $\sqrt s= 57$ TeV, we conclude that: i) the total $pp$ cross section is $\sigma_{\rm tot}=133.4\pm 1.6$ mb,
ii) $\spai$, the PAO p-air inelastic production cross section , after correction for a 19\% helium contamination, is given by $\spai= 482\pm 30$ mb
iii) our prediction for the $pp$ total cross section, $\sigma_{\rm tot}$, taken from Ref. \cite{blackdisk}, yields a $pp$ inelastic cross section $\sigma_{\rm inel}= 92.9\pm 1.6$ mb, which is consistent with the PAO \cite{Augerinelastic} result of $\sigma_{\rm inel}= 90\pm 7\  ({\rm stat.}) \pm^9_{11}\  ({\rm syst.}) \pm 1.5 {\rm \  (Glaub.})$ mb. 

{\em Acknowledgments}. M.M.B. would like to thank the Aspen Center for Physics, supported in part by NSF Grant No. 1066293, for its
 hospitality during the writing of this manuscript. He would like to thank his colleague Francis Halzen for invaluable discussions and aid during the preparation of this manuscript.


\begin{thebibliography}{999}
\bibitem{blackdisk} 
M. M. Block and F. Halzen, arXiv:1109:2041, 2011. 


\bibitem{Auger} 
The Pierre Auger Collaboration,  R. Ulrich, part2, "Estimate of the proton-air cross section", arXiv:1107.4804 [astro-phys.HE],  (2011).
\bibitem{Augerinelastic}
The Pierre Auger Collaboration, M. Mostaf\'a, XXXI Physics in Collision Conference, Vancouver, Sept. 1, 2011.    
\bibitem{Blockcr} 
M. M. Block, Phys. Rev D{\bf 76}, 111503, 2007.
\bibitem{BH} 
M. M. Block and F. Halzen, Phys. Rev. D{\bf 72}, 036006, 2005.
\bibitem{sieve}
M. M. Block, Nucl. Instrum. Methods A {\bf 556}, 308, 2006.
\bibitem{blockanalyticity}
M. M. Block, Eur. Phys J. C{\bf 47}, 697, 2006.
\bibitem{blockhalzenstanev} 
M. M. Block, F. Halzen and T. Stanev, Phys. Rev. Lett. {\bf 83}, 4926, 1999; Phys. Rev. D{\bf 62} 77501, 2000.
\bibitem{froissart} 
M. Froissart, Phys. Rev. {\bf 123}, 1053, 1961. 
%
\end{thebibliography}
\end{document}